\documentclass[prd,aps,showpacs,preprint,nofootinbib,floatfix,superscriptaddress, showkeys]{revtex4-1}
\usepackage{graphics}
\usepackage{epsfig}
\usepackage{latexsym}
\usepackage{colordvi}
\usepackage{amsmath}
\usepackage{amssymb}
\begin{document}
\preprint{\parbox[b]{1in}{ \hbox{\tt PNUTP-17/A05} }}
\preprint{\parbox[b]{1in}{  \hbox{\tt APCTP Pre2017-010} }}

\title{Clockwork graviton contributions to muon $g-2$}


\author{Deog Ki Hong}
\email[]{dkhong@pusan.ac.kr}
\affiliation{Department of
Physics,   Pusan National University,
             Busan 46241, Korea} 
\author{Du Hwan Kim}
\email[]{kdh314@pusan.ac.kr}
\affiliation{Department of
Physics,   Pusan National University,
             Busan 46241, Korea} 
\affiliation{Center for Theoretical Physics of the Universe, Institute for Basic Science,
   Daejeon, 34051, Korea}
\author{Chang Sub Shin}
\email[]{csshin@ibs.re.kr}
\affiliation{Center for Theoretical Physics of the Universe, Institute for Basic Science,
   Daejeon, 34051, Korea}
\affiliation{Asia Pacific Center for Theoretical Physics, Pohang 37673, Korea}
\affiliation{Department of Physics, Postech, Pohang 37673, Korea}

\date{\today}

\begin{abstract}
The clockwork mechanism for gravity introduces a tower of massive graviton modes, ``{\it clockwork gravitons},'' with a very compressed mass spectrum, whose interaction strengths are much stronger than that of massless gravitons. In this work, we compute the lowest order contributions of the clockwork gravitons to the anomalous magnetic moment, $g-2$, of muon in the context of extra dimensional model with a five dimensional Planck mass, $M_5$. 
We find that the total contributions are rather insensitive to the detailed model parameters, and  determined mostly by the value of $M_5$. In order to account for the current muon $g-2$ anomaly, $M_5$ should be around $0.2~{\rm TeV}$, 
and the size of the extra dimension has to be quite large, $l_5 \gtrsim 10^{-7}\,$m.  For $M_5\gtrsim1~{\rm TeV}$, the clockwork graviton contributions are too small  to explain the current muon $g-2$ anomaly. 
We also compare the clockwork graviton  contributions with other extra dimension models such as Randall-Sundrum models or large extra dimension models. We find that the leading contributions in the small curvature limit are universal, but the cutoff-independent subleading contributions vary for different background geometries and the clockwork geometry gives the smallest subleading contributions. 
	
\end{abstract}


\maketitle

\subsection{Introduction}
After the Higgs boson was discovered to complete the standard model (SM) of particle physics, there has been an intense search for new particles at the large hadron collider (LHC) that probes ${\rm TeV}$ energy scales. At LHC Run 2 the mass limit of new particles  has been pushed up above $1~{\rm TeV}$~\cite{Khachatryan:2016yec,ATLAS:2016eeo,Aaboud:2017efa}, putting most models for physics beyond SM (BSM) in great tension with their naturalness criterion, advocated by 't Hooft~\cite{'tHooft:1979bh}. 

Recently an interesting mechanism, called clockwork (CW), is proposed to generate naturally an exponential hierarchy for a given theory with multicomponents of fields~\cite{Choi:2014rja,Choi:2015fiu,Kaplan:2015fuy}. 
Giudice and McCullough then proposed a clockwork solution to the electroweak hierarchy problem~\cite{Giudice:2016yja}, which exhibits rather rich structure and phenomenology~\cite{Giudice:2017fmj}.
The clockwork scenario addresses, similarly to other extra dimensional scenarios, the hierarchy problem by assuming that the fundamental scale of the theory is not much higher than the electroweak scale. 
The simplest clockwork model can be constructed with a set of 4D theories at $ N+1$ sites in a theory space with asymmetric couplings of link fields between nearby sites so that the zero mode of link fields is highly localized at a single site, while all SM particles reside at a site which has the least overlap with the zero mode to suppress its coupling to SM particles. If one identifies this zero mode as the 4D massless graviton\footnote{The clockwork symmetry between two neighboring sites is shown to be respected only for the abelian case~\cite{Craig:2017cda}. But, here we take the effective field theory approach, as emphasized in~\cite{Giudice:2017suc}, for the clockwork gravity that is nothing but the discretized linear-dilaton model~\cite{Antoniadis:2011qw,Baryakhtar:2012wj}.}, the clockwork setup solves the naturalness problem associated with the weak scale and becomes in the large $N$ limit the linear dilaton model~\cite{Antoniadis:2011qw,Baryakhtar:2012wj} of 5D little string theory~\cite{Aharony:1998ub,Aharony:2004xn}. The clockwork theory  then predicts an infinite tower of massive Kaluza-Klein (KK) gravitons, with unique spectrum, that couple to SM particles in a specific way. Especially the low-lying states of clockwork KK gravitons exhibit rather interesting signatures at colliders, compared to other models of extra dimensions, as studied in detail~\cite{Giudice:2017fmj}.
In this paper we study the contributions of the clockwork gravitons to the anomalous magnetic moments of muon and constrain the parameters of the clockwork gravitons, which will be complementary to collider searches. We find that the intrinsic scale, $M_5$,  of  the clockwork graviton has to be around $0.2~{\rm TeV}$ or higher to be compatible with the current muon $g-2$ anomaly. 
 
It is well known that the standard model estimation of the anomalous magnetic moment of muon has quite a significant deviation from  the experiments, which thus provides interesting constraints for new physics, if it were to explain the deviation~\footnote{The muon $g-2$ would also provide interesting constraints on new physics even if there were not any deviation.}. 
The current deviation between the experimental value, obtained at the Brookhaven National Laboratory (BNL)~\cite{Bennett:2006fi}, and the SM estimate, based on $e^{+}e^{-}$ hadronic cross sections~\cite{Jegerlehner:2009ry}, is found to be 
\begin{equation}
\Delta a_{\mu}=a_{\mu}^{\rm exp}- a_{\mu}^{\rm th}=(290\pm90)\times10^{-11}\,,
\end{equation}
where $a_{\mu}\equiv \left(g-2\right)/2$ and $g$ is the gyromagnetic ratio of the magnetic moment of muon. 
The current deviation corresponds to about $3.2~\sigma$.  An improved muon $(g-2)$ experiment at the Fermilab is about to take data, aiming to achieve a precision of $0.14$ ppm~\cite{Miller:2007kk,Gray:2015qna}, which will then move the current deviation, if persistent, to more than 5~$\sigma$.

\subsection{Massive graviton constributions to muon $g-2$}
While a massless spin 2 particle, that couples to the energy-momentum tensor, necessarily leads at low energy to Einstein's general relativity,  a consistent description of massive gravitons, respecting the general coordinate invariance, has been found only recently~\cite{deRham:2014zqa}. 
To describe a (massive) graviton in a flat spacetime, we write the metric as 
\begin{equation}
g_{\mu\nu}=\eta_{\mu\nu}+2\kappa\,h_{\mu\nu}\,
\end{equation}
where $\eta_{\mu\nu}$ is the Minkowski metric for the flat spacetime and $h_{\mu\nu}$ is the graviton field with coupling $\kappa\equiv\sqrt{8\pi G}$ for Einstein gravity with Newton's gravitational constant $G$.
Under a general coordinate transformation, $x^{\mu}\mapsto x^{\mu}+\xi^{\mu}(x)$, the graviton field transforms as
\begin{equation}
h_{\mu\nu}\mapsto h_{\mu\nu}-\frac{1}{2\kappa}\left(\partial_{\mu}\xi_{\nu}+\partial_{\nu}\xi_{\mu}\right)\,.
\end{equation}
Fixing the above gauge degrees of freedom, the massive graviton propagator of mass $M$ in $D$ dimensional spacetime becomes 
\begin{equation}
\int_xe^{ip\cdot x}\left<0\right|T\left\{h_{\mu\nu}(x)h_{\alpha\beta}(0)\right\}\left|0\right>=\frac{i}{2}\frac{{\tilde \eta}^{\mu\alpha}
{\tilde \eta}^{\nu\beta}+{\tilde \eta}^{\mu\beta}{\tilde \eta}^{\nu\alpha}-\beta{\tilde \eta}^{\mu\nu}{\tilde \eta}^{\alpha\beta}}{p^2-M^2+i\epsilon}\,,
\end{equation}
where ${\tilde \eta}^{\mu\nu}=\eta_{\mu\nu}-p^{\mu}p^{\nu}/M^2$ and $\beta=\frac{2}{D-1}$. (For massless gravitons, ${\tilde \eta}^{\mu\nu}=\eta^{\mu\nu}-p^{\mu}p^{\nu}/p^2$ and $\beta=\frac{2}{D-2}$. Here we consider $D=4$ only.) The gravitational interactions are given at the linear level as
\begin{equation}
{\cal L}_{\rm int}=-{\kappa}\, h_{\mu\nu}T^{\mu\nu}\,,
\end{equation}
where $T^{\mu\nu}\equiv\frac{-2}{\sqrt{-g}}\frac{\delta S}{\delta g_{\mu\nu}}$  with $S$ being the SM action is a symmetric and conserved energy-momentum tensor of SM that sources (massive) gravitons.

\begin{figure}[h!]
\vskip 0.1in
    \centering
    \begin{minipage}{0.15\textwidth}
        \centering
        \includegraphics[width=1\linewidth, height=0.12\textheight]{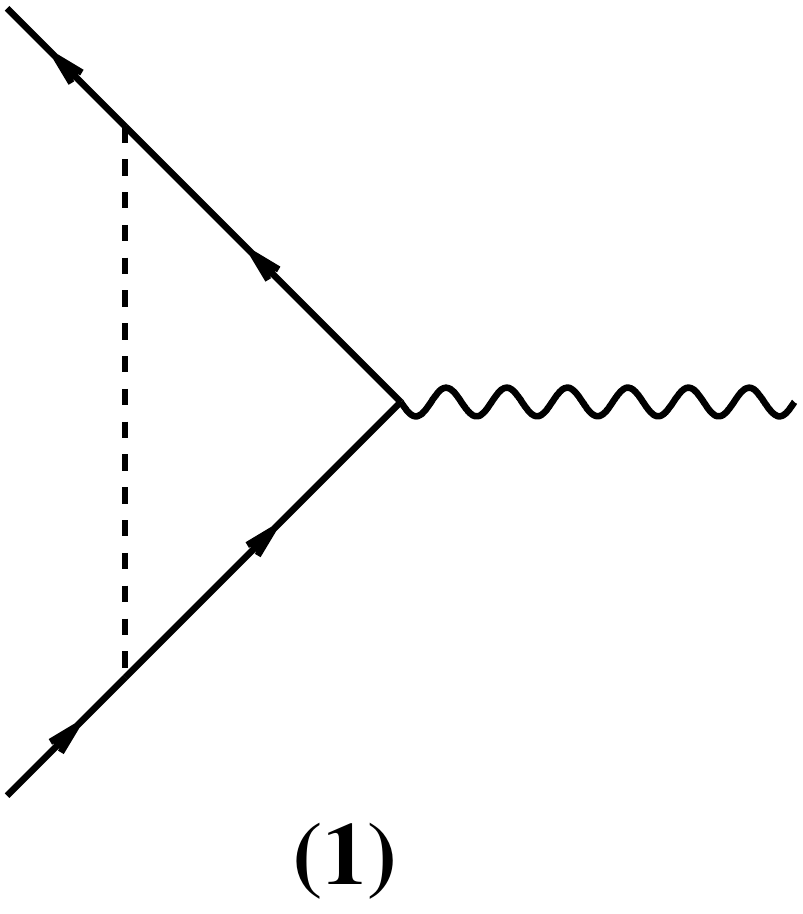}
    \end{minipage}%
   \hspace{.2in}
      \centering
    \begin{minipage}{0.15\textwidth}
        \centering
        \includegraphics[width=1\linewidth, height=0.12\textheight]{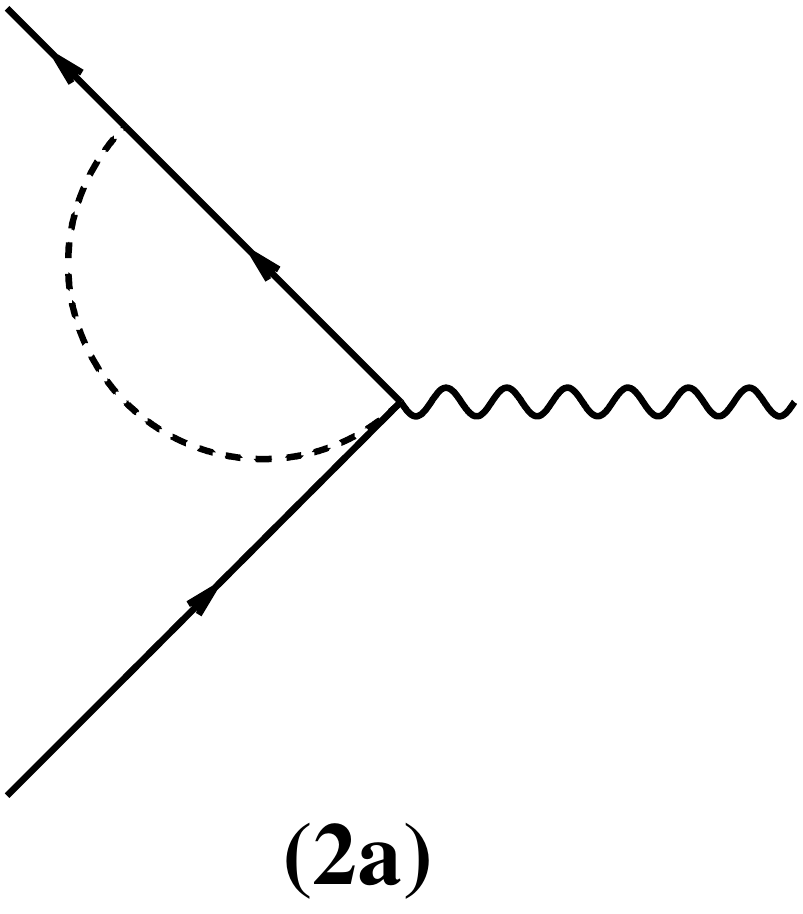}
    \end{minipage}%
     \hspace{.2in}
  \centering
    \begin{minipage}{0.15\textwidth}
        \centering
        \includegraphics[width=1\linewidth, height=0.12\textheight]{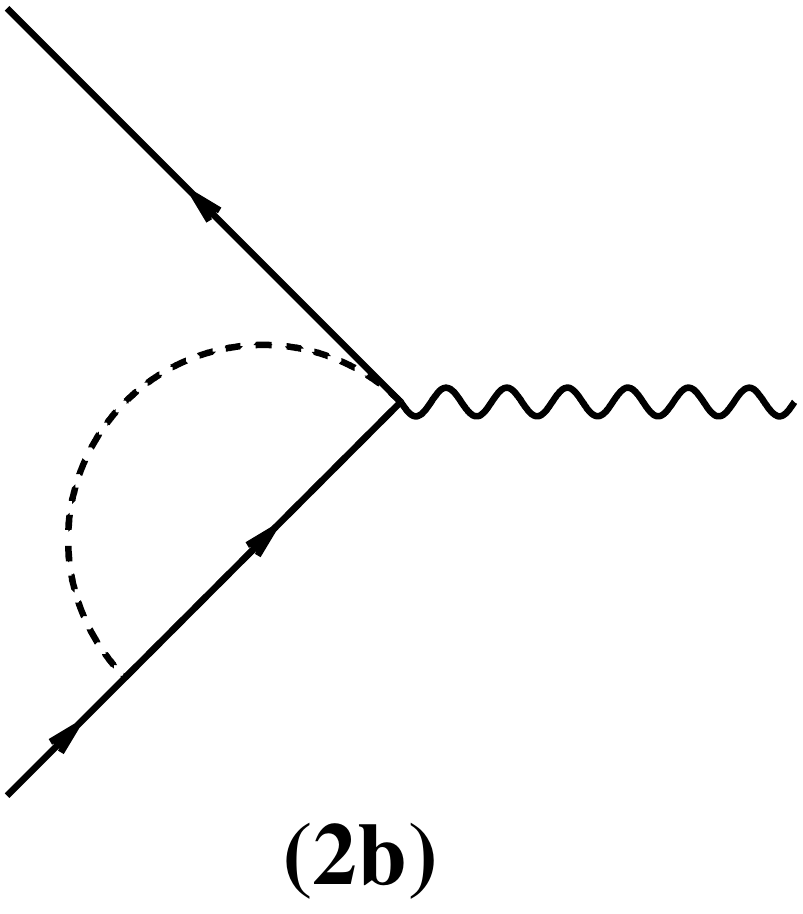}
    \end{minipage}%
     \hspace{.2in}
  \centering
    \begin{minipage}{0.15\textwidth}
        \centering
        \includegraphics[width=1\linewidth, height=0.12\textheight]{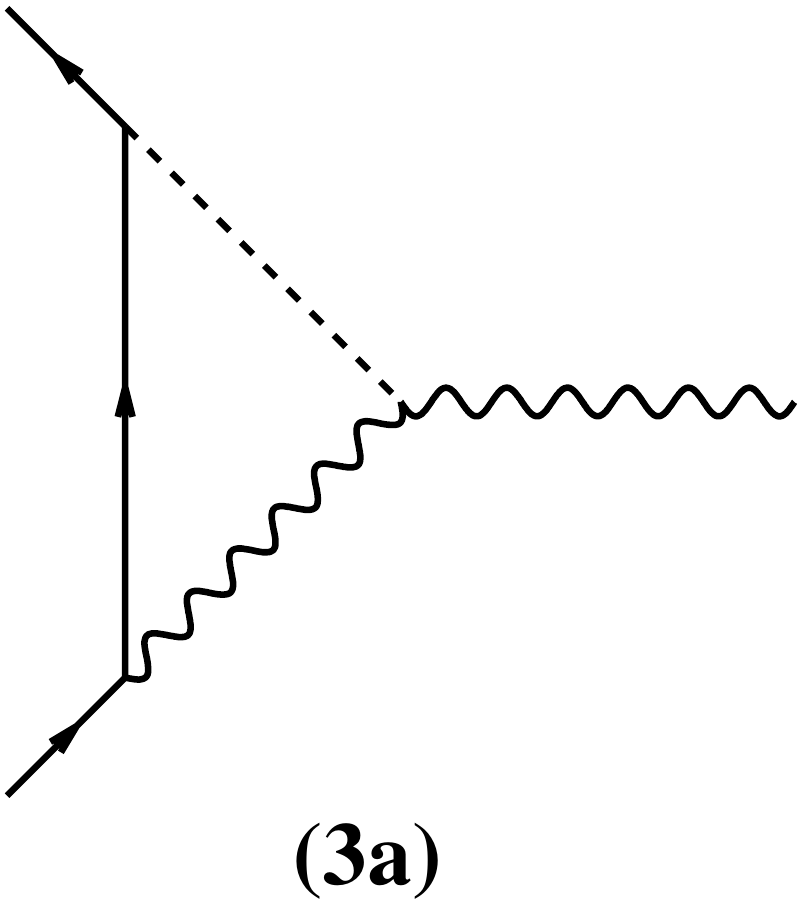}
    \end{minipage}%
     \hspace{.2in}
  \centering
    \begin{minipage}{0.15\textwidth}
        \centering
        \includegraphics[width=1\linewidth, height=0.12\textheight]{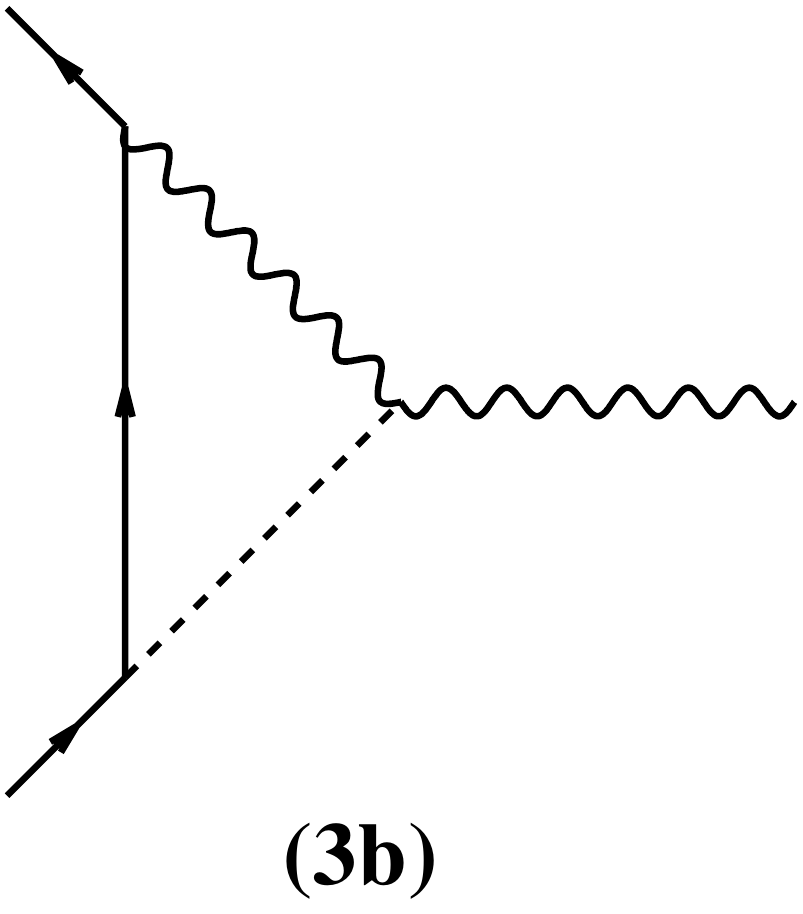}
    \end{minipage}%
  \caption{
Graviton contributions at one-loop to the anomalous magnetic moment of muon. The dashed lines denote gravitons and the curly lines are photons.
}
  \label{fig1}   
\end{figure}

\begin{figure}[h!]
  \centering
    \includegraphics[width=0.5\textwidth]{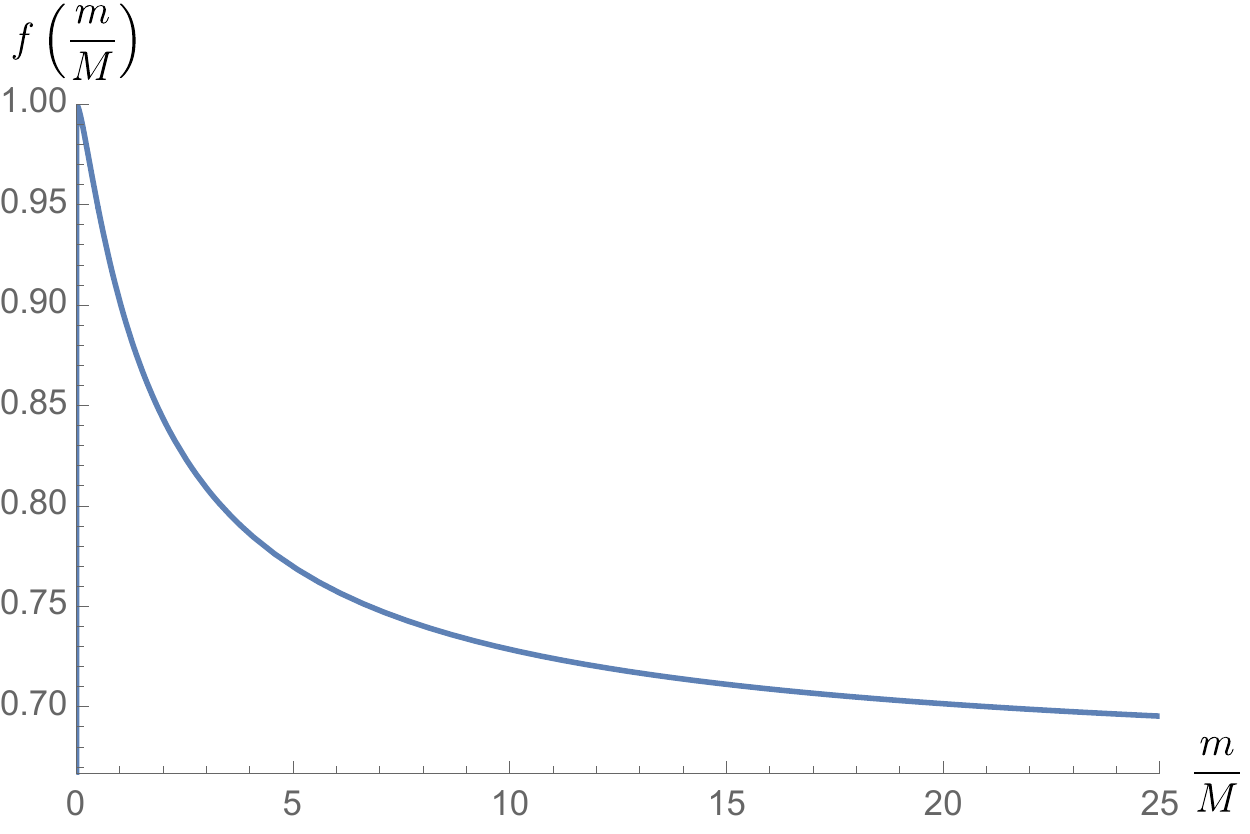}
     \caption{The massive graviton contributions in the unit of $a_{\mu}^{\rm massive}(M\to\infty)$.}
      \label{fig2}
\end{figure}

The one-loop  contributions of massless graviton to muon anomalous magnetic moment was calculated by Berendes and Gastmans~\cite{Berends:1974tr} and that of massive graviton was calculated in~\cite{Graesser:1999yg}. Both are found to be finite. We briefly discuss the single graviton contributions first.  There are 5 diagrams that contribute at one-loop to muon $g-2$, shown in Fig.~\ref{fig1}.
All the five diagrams are ultraviolet (UV) divergent and hence need to be regularized. But, the sum of all five diagrams for muon $g-2$ turns out to be finite. For massless graviton, one finds 
\begin{equation}
a_{\mu}^{\rm gr}=\frac{7}{32\pi^2}\kappa^2m^2\,,
\end{equation}
where $m$ is the mass of muon~\cite{Berends:1974tr}. 
For the one-loop contribution of massive gravitons of mass $M$ to muon $g-2$, we find  
\begin{equation}
a_{\mu}^{\rm massive}=\frac{5}{16\pi^2}\kappa^2m^2f\left(\frac{m}{M}\right)\,,
\end{equation}
where $f(x)$ is a monotonically decreasing function from $1$ to $2/3$ as $x$ increases from 0 to $\infty$ (See Fig.~\ref{fig2}), which agrees with the previous result in the integral form in~\cite{Graesser:1999yg}~\footnote{There is a minor typo in Eq.~(16) of \cite{Graesser:1999yg} that the function in the integrand should be $R$ instead of $R/L$.}.
We note that the massive graviton does not decouple in the loop corrections to muon $g-2$, as its mass goes to infinity, $M/m\to\infty$, since the gravity is non-renormalizable~\footnote{The graviton coupling to muon grows in energy as $\kappa^2E^2$ and the massive graviton contribution to the muon $g-2$ is thus not suppressed. The unitarity is also expected to be violated at energy $E\sim\kappa^{-1}\equiv M_{P}\,(=2.4\times 10^{18}~{\rm GeV})$. But, because the gravity is nonlinear, the theory necessarily has a UV cutoff below the Planck scale, $M_P$~\cite{ArkaniHamed:2002sp}.}. 
In the large mass limit, $M\gg m$,  
\begin{equation}
a_{\mu}^{\rm massive}=\frac{5}{16\pi^2}\kappa^2m^2\left[1+\left\{\frac{1}{3}\ln\left(\frac{m}{M}\right)+\frac{11}{72}\right\}\frac{m^2}{M^2}+\cdots\right]\,,
\label{massive}
\end{equation}
where the ellipsis denotes the higher order terms in $m/M$. 
The massless limit of massive graviton is known to be discontinuous due to the non-decoupling of the longitudinal mode~\cite{vanDam:1970vg}. For the muon anomalous magnetic moment one finds a following discontinuity~\cite{Graesser:1999yg}:
\begin{equation}
a_{\mu}^{\rm massive}(M\to0)=a_{\mu}^{\rm gr}+\frac{\kappa^2m^2}{48\pi^2}\left(1-\frac{D-1}{D-2}\right)\,, 
\end{equation}
where $D=4$ is the dimension of the spacetime in which massive gravitons propagate.

\subsection{Clockwork gravitons and other extra dimension models}
If gravitons propagate in extra dimensions as well as in 4D spacetime where SM particles reside, either 
 in a continuous extra dimension such as the large extra dimension (LED) model~\cite{ArkaniHamed:1998rs},  and  the Randall-Sundrum (RS) model~\cite{Randall:1999ee}, or 
in a discrete extra dimension as the deconstructed gravity~\cite{ArkaniHamed:2002sp} and  the clockwork (CW) gravity~\cite{Giudice:2016yja}, there will be a (infinite) tower of massive gravitons, $h^{(n)}_{\mu\nu}$, that interact with 4D SM particles with specific couplings,  
\begin{equation}
{\cal L}_{\rm int}=-\sum_{n=1}\frac{1}{\Lambda_n}h_{\mu\nu}^{(n)}T^{\mu\nu}.
\label{massive_int}
\end{equation}
The mass of the $n$-th graviton $M_{(n)}$ and  the coupling $1/\Lambda_n$ are the function of the model parameters and its intrinsic scale $M_5$ that describes the extra dimension, discussed below. In this paper, we are mostly interested in the clockwork case. However it turns out that 
the leading contribution to the muon $g-2$ is quite independent of detailed model parameters, and 
is only the function of $m$ and $M_5$. Let us start from the CW case first, and we will discuss later the cases of RS and LED backgrounds.

In the clockwork theory the massive gravitons, $h_{\mu\nu}^{(n)}$, have mass, given as 
\begin{equation}
M_{(n)}^2=k^2+\frac{n^2}{R^2}+{\cal O}\left(\frac{1}{N}\right),\quad n=1,\cdots,N\,,
\label{mass}
\end{equation}
where $k$, $R$ correspond respectively to the warped factor or the clockwork spring and the radius of the extra dimension, orbifolded by $Z_2$. 
In the continuum limit ($N\to\infty$) the clockwork geometry becomes, having the SM particles localized at the $y=0$ brane, 
\begin{equation}
ds^2=e^{\frac{4k\left|y\right|}{3}}\left(dx_{\mu}dx^{\mu}-dy^2\right)\,,
\end{equation}
which is nothing but the linear dilaton model studied in~\cite{Antoniadis:2011qw}. Being a 5D theory, the continuum clockwork theory has an intrinsic 5D Planck scale $M_5$, which is an additional parameter of the clockwork theory,  in addition to the radius $R$ and  the warped factor $k$ or the 5D cosmological constant $-2k^2$.
Upon the Kaluza-Klein (KK) reduction of the 5D clockwork gravity, one finds
the clockwork gravitons' couplings to the SM particles at the linear level~\cite{Giudice:2016yja}, given as the inverse of 
\begin{equation}
\Lambda_n=\sqrt{M_5^3\pi R\left(1+\frac{k^2R^2}{n^2}\right)}\,
\end{equation}
and the effective 4D Planck mass $M_P$ or the effective coupling of the massless graviton,  $\kappa=1/M_P$, is defined as 
\begin{equation}
M_P^2=\frac{M_5^3}{k}\left(e^{2k\pi R}-1\right)\,.
\label{planck}
\end{equation}

The $n$-th clockwork graviton contribution to the muon $g-2$ at one-loop is from Eqs.~(\ref{massive}) and (\ref{massive_int})
\begin{equation}
a_{\mu}^{(n)}=\frac{5}{16\pi^2}\frac{m^2}{\Lambda_n^2}\left[1+{\cal O}\left(\frac{m^2}{M_n^2}\right)\right]\,.
\end{equation}
Since massive gravitons do not decouple to the muon anomalous magnetic moment in the large mass limit ($M_{(n)}\gg m$), one might assume all towers of gravitons do contribute to the muon $g-2$. However, since the effective description of massive gravitons in 4D will breakdown at very short distances, smaller than the UV cutoff, $\Lambda_{\rm cut}$, of the clockwork theory,  only finite number of gravitons are relevant in the 4D effective theory. 
The relevant gravitons should have therefore mass $M_{(n)}\lesssim \Lambda_{\rm cut}$ or the highest level $n_c$ that massive gravitons can be excited in the 4D effective theory should be from Eq.~(\ref{mass})
\begin{equation}
n_c=\sqrt{(\Lambda_{\rm cut}R)^2-(kR)^2}\,.
\end{equation}

Since the choice of the cutoff $\Lambda_{\rm cut}$ depends on the regularization scheme of the 5D effective theory~\cite{Contino:2001nj}, we parametrize the regularization scheme dependence by a constant $\alpha$ as
\begin{equation}
\Lambda_{\rm cut}=\alpha M_5\,,
\end{equation}
where $\alpha$ should be positive and expected to be ${\cal O}(1)$ by the naive dimensional analysis but not predictable in the 5D effective theory.
We will keep this parameter in our discussion in order to understand its physical meaning, and then take $\alpha=1$ just to estimate some numerical values.
The 5D Planck mass $M_5$ of the clockwork theory also sets the upper bound of the curvature of 5D clockwork geometry. For the clockwork to work the warped factor should be smaller than the intrinsic scale, $k\ll M_5$, and we have $n_c\gg 1$ for a given $kR$. 
Furthermore, since we are considering only the linear terms for the graviton interactions, our approximation will break down when massive gravitons couple strongly,  where the effects of  UV completed quantum gravity are important.  Therefore, there should be an upper bound for the highest level,  $n_c\le n_{*}$, the massive gravitons of the 4D effective theory can reach. Namely  at the upper bound $n=n_{*}$, the graviton mass is of the order of its effective Planck scale, $\Lambda_{n_*}\approx M_{(n_{*})}$. Approximately we then have 
\begin{equation}
M_5^3\pi R\left(1+\frac{k^2R^2}{n_*^2}\right)=k^2+\frac{n_*^2}{R^2}\,.
\end{equation}
The maximum upper bound for graviton mass is therefore given by  $n_*=M_5R\sqrt{M_5\pi R}$, as three parameters of the clockwork theory $k,R$ and $M_5$ are related to each other by Eq.~(\ref{planck}).
The number of allowed KK levels, $n_c$, or the hierarchies between $M_{(1)}\simeq k$, $1/R$ and $M_5$ 
are very sensitive to the value of $kR$, as shown in Tables~\ref{tab1} and \ref{tab2}, which leads 
to very different collider phenomenology. 
However the muon $g-2$ is rather insensitive to such model parameters for $k\ll M_5$.
Since the more the allowed KK levels are, the weaker their couplings, two effects cancel with each other.
\begin{table}[ht]
\caption{The maximum Kaluza-Klein graviton level $n_c$ in the clockwork geometry with $k=10/R$, $n_*$ in the 4D effective theory, 
and the contribution to the muon $g-2$ for different $M_5$ with $\alpha=1$.}
\centering
\begin{tabular}{c| c c c c}
\hline\hline
$M_5~(\rm TeV)$ & $M_5R$ & $n_c$ & $n_*$ & $a_{\mu}^{\rm CW}\times 10^{10}$ \\ [0.5ex] 
\hline
0.5&\ $1.2\times 10^5$ &\ $1.2\times 10^5$ \ &$7.4\times 10^7$ & 4.5 \\
1&\ $3.1\times 10^4$ &\ $3.0\times 10^4$ \ &$9\times 10^6$  & 1.1 \\
5&\ $1.2\times 10^3$ &\ $1.2\times 10^3$ \ & $7.3\times 10^4$  &  0.045  \\
10 &\ 306 &\ 305 \ &\ $9.5\times 10^3$ &  0.011    \\
50 &\ 12 &\ 7 \ & 74 & $4.0\times 10^{-5}$ \\ [1ex]
\hline
\end{tabular}
\label{tab1}
\end{table}
\begin{table}[ht]
\caption{Same as Table \ref{tab1}, but fixing $k$ as $k = 0.1 M_5$.}
\centering
\begin{tabular}{c | c c c c}
\hline\hline
$M_5~(\rm TeV)$ & $kR$ & $n_c$ & $n_*$ & $a_{\mu}^{\rm CW}\times 10^{10}$ \\ [0.5ex] 
\hline
0.5&\ 11.1 &\ 111 \ &$2.1\times 10^3$ & 3.9 \\
1&\ 10.9 &\ 109 \ &$2.1\times 10^3$  & 1.0 \\
5&\ 10.4&\ 102 \ &$1.9\times 10^3$  & 0.038 \\
10 &\ 10.2 &\ 99 \ & $1.8\times 10^3$ & 0.010  \\
50 &\  9.7&\ 93 \ & $1.7\times 10^3$ &\ $3.9\times 10^{-4}$ \\ [1ex]
\hline
\end{tabular}
\label{tab2}
\end{table}
Summing up the contributions of the clockwork gravitons up to the $n_c$-th level, we get 
\begin{equation}
a_{\mu}^{\rm CW}\simeq\sum_{n=1}^{n_c}a_{\mu}^{(n)}=\frac{5}{16\pi^3}\left(\frac{m}{M_5}\right)^2\left[\alpha
\sqrt{1-\left(\frac{k}{\alpha M_5}\right)^2}-\frac{k}{M_5}\sum_{n=1}^{n_c}\frac{kR}{n^2+(kR)^2}\right]\,, 
\end{equation}
which becomes for $n_c\gg1$ or $M_5\gg k$ 
\begin{equation}
a_{\mu}^{\rm CW}\simeq \frac{5}{16\pi^3}\left(\frac{m}{M_5}\right)^2\left[\alpha-\frac{k}{M_5}\left\{-\frac{1}{2kR}+\frac{\pi}{2}\coth(\pi kR)\right\}+{\cal O}\left(\frac{k^2}{M_5^2}\right)\right]\,.
\end{equation}

We note that  the contribution of clockwork gravitons could be divided by the regularization scheme-dependent part, 
\begin{equation}
\Delta_{(1)} a_\mu^{\rm CW} \simeq\frac{5 \alpha }{16\pi^3}\left(\frac{m}{M_5}\right)^2 , 
\end{equation}
which is independent of detailed graviton spectrum, and the scheme-independent part
\begin{equation}
\Delta_{(2)} a_\mu^{\rm CW} \simeq  - \frac{5}{32\pi^2}\frac{k}{M_5}\left(\frac{m}{M_5}\right)^2 ,
\end{equation} which depends on the 5D geometry. 
The fact that $\Delta_{(1)}a_\mu^{\rm CW}$ is independent of background geometry  is easy to understand. 
In a 5D theory, this term represents a linear divergence that is dominated by large graviton momentum, $p_5$, along the fifth dimension.
Such a graviton of large momentum would not see the background geometry.  Also 
this term could be absorbed by a local counter term for the graviton-loop contribution such as 
\begin{equation}
{\cal L}_{\rm c.t.} \sim \frac{\lambda}{M_5^2}( \bar \psi \gamma^{\mu\nu} \gamma^\rho D_\rho \psi) F_{\mu\nu}, 
\end{equation} which is consistent with symmetries of theory and independent of the background geometry. 
On the other hand, the contribution $\Delta_{(2)}a_\mu^{\rm CW}$ is UV finite and scheme-independent. 
Therefore it can be considered as the contribution that represents a genuine feature of CW geometry.  
In order to see that $\Delta_{(2)} a_\mu^{\rm CW}$ is scheme-independent, we may regularize the KK sum by different methods. For example, using
\begin{align}
\sum_{n=1}^{\infty} \frac{n^2}{n^2 + (kR)^2}  
&= \left[\frac{\partial^2}{\partial \epsilon^2} \sum_{n=1}^{\infty} \frac{ e^{- \epsilon n}}{n^2 + (kR)^2}\right]_{\epsilon \to 0^+}\nonumber\\
&= \frac{1}{\epsilon} +\frac{1}{2} +\frac{i kR}{2}\left[\frac{\Gamma'( i k R)}{\Gamma( i k R)}- {\rm h.c.}\right]\nonumber\\
&\approx \frac{1}{\epsilon} - \frac{\pi k R}{2} \quad{\rm for}\ kR\gg 1, 
\end{align}  
where ${\rm h.c.}$ denotes the Hermitian conjugate and $\Gamma(x)$, $\Gamma^{\prime}(x)$ the gamma function and its derivative, respectively, 
we get the same finite contribution for $\Delta_{(2)}a_\mu^{\rm CW}$.

\begin{figure}[h!]
  \centering
    \includegraphics[width=0.6\textwidth]{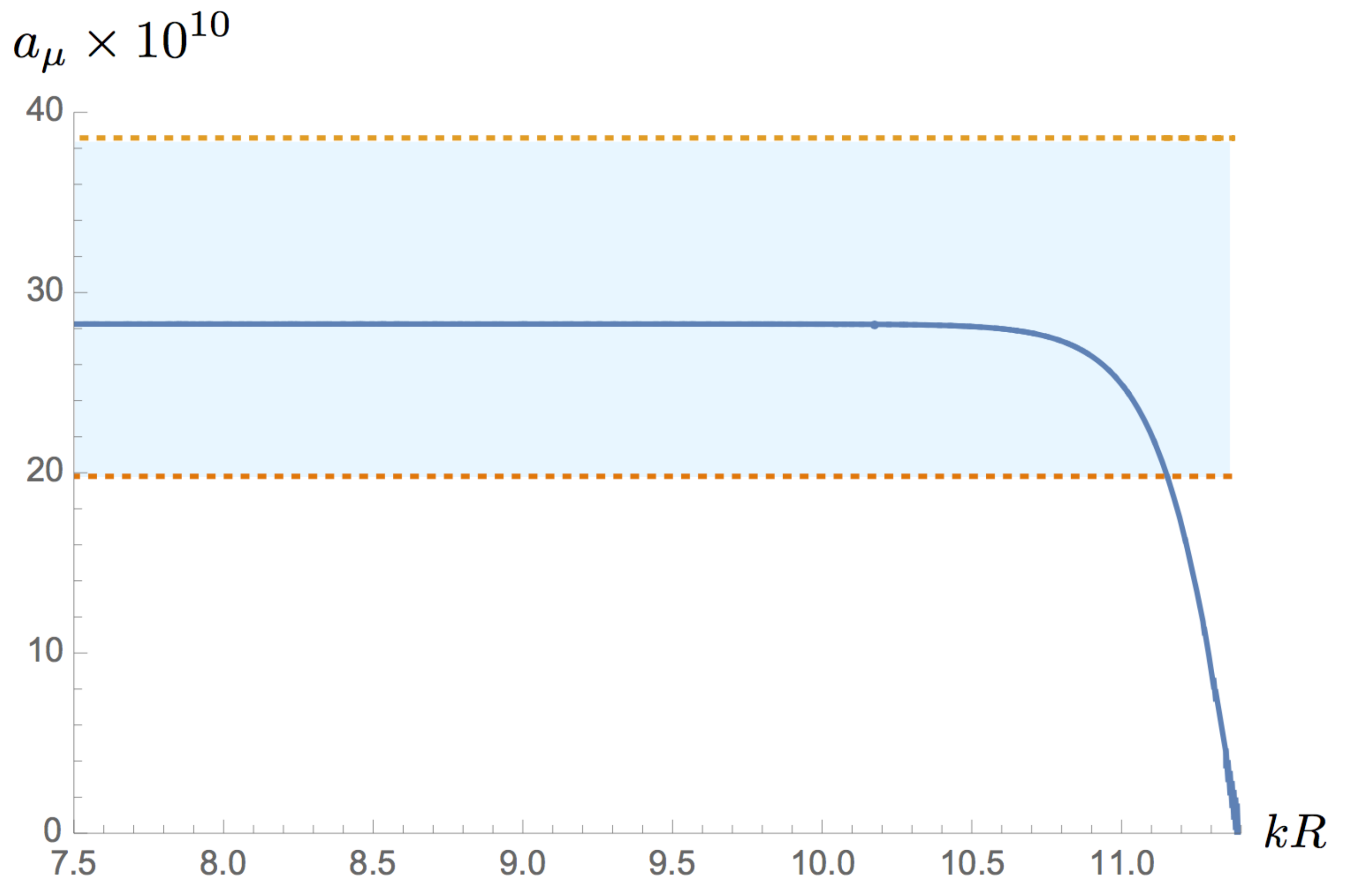}  
      \caption{The clockwork graviton contributions, shown in solid line, to the muon $g-2$ for $M_5=0.2~{\rm TeV}$ and $kR\in[7.5,11.3]$. The $1\sigma$ band of the muon $g-2$ anomaly lies in the shaded region between two dashed lines. }
      \label{fig3}
\end{figure}

In the absence of a computation in a UV complete theory, however, it is still useful to calculate  whole clockwork graviton contributions, taking $\alpha=1$, although a UV complete theory is needed to determine $\alpha$.
Taking $kR\simeq10$, we get respectively for $M_5=0.5,1,10,50~{\rm TeV}$ as 
\begin{equation}
a_{\mu}^{\rm CW}\times 10^{10}\simeq 4.5,\,1.1,\,0.045,\,0.01,\,4.5\times 10^{-4}\,.
\end{equation}
For $M_5=0.2~{\rm TeV}$, we find that the CW theory does explain the muon $g-2$ anomaly for any value of $k$ up to its maximum value $k_{\rm max}\sim M_5$, which corresponds to $k_{\rm max}R\sim11$ or $n_c\sim {\cal O}(1)$ (see Fig.~\ref{fig3}).
\begin{figure}[h!]
  \centering
      \includegraphics[width=0.45\textwidth]{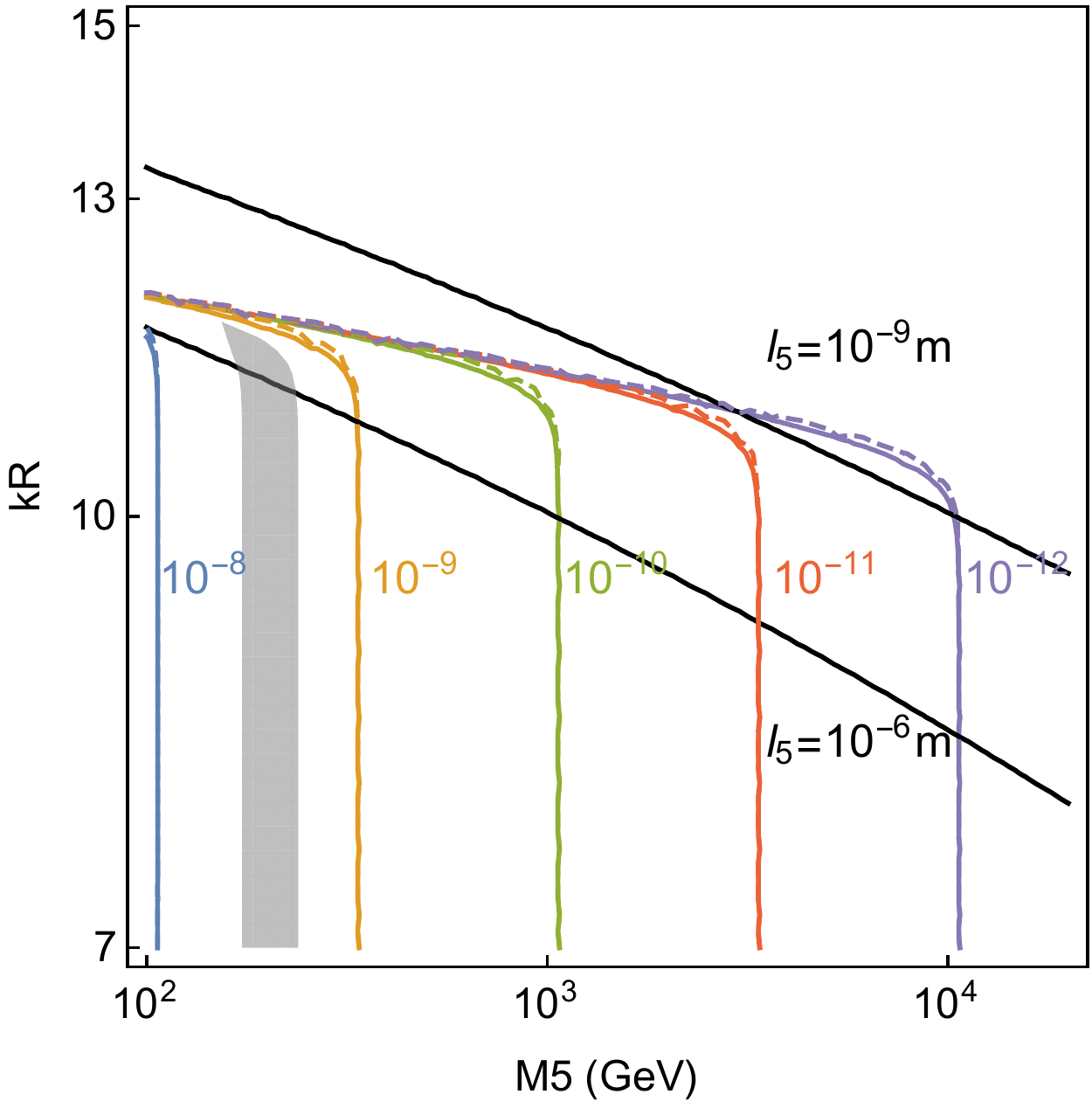}\  
     \includegraphics[width=0.46\textwidth]{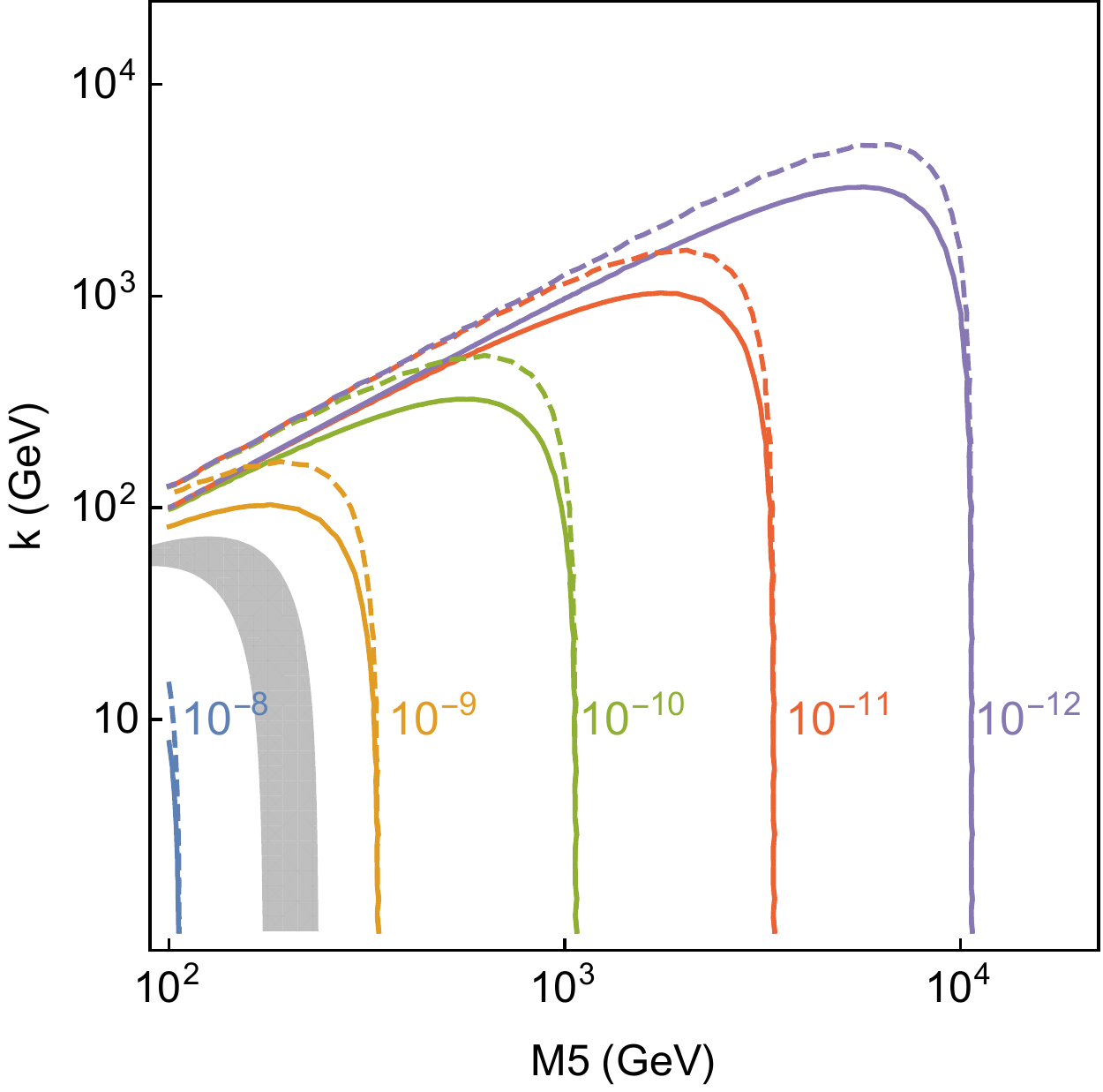}
     \caption{The  contribution of the KK gravitons to the muon $g-2$ for $\Delta a_\mu = 10^{-8},\cdots, 10^{-12}$ in the $M_5-kR$ (left) and $M_5-k$ (right) plots. The contributions from the clockwork gravitons are shown by the solid lines. Here we also show the contributions 
with the Randall-Sundrum background, Eq.~(\ref{muong-2RS}) by dashed lines. 
The gray region is the $1\sigma$ band of the current muon $g-2$ anomaly~\cite{Bennett:2006fi}. 
In the left panel, the size of extra dimension, $l_5$  (Eq.~(\ref{extrasize})) is shown by black lines for $l_5=10^{-6}\,$m and $10^{-9}\,$m.}
      \label{fig4}
\end{figure}
The KK graviton contributions for generic values of $k$ and $M_5$ are plotted in Fig.~\ref{fig4}~\footnote{ Note that in this plot we did not impose the constraints on the low values of $M_5$ from other experiments just to see the parameter dependence of $a_\mu$ more clearly. However, the muon $g-2$ anomaly alone already excludes the clockwork theory with $M_5<0.2~{\rm TeV}$, unless there are negative contributions from other new particles.}.
In Fig. 4, we also show that there is a lower bound for the size of extra dimension $l_5$, defined as
\begin{equation}
l_5 \equiv\int_0^{\pi R} dy \sqrt{ -g_{55}} = \frac{3(e^{\frac{2}{3} k\pi R} - 1)}{2k}, 
\label{extrasize}
\end{equation}
to explain the muon $g-2$ anomaly. 
If $l_5$ is smaller than $10^{-7}$m, the CW gravitons contribution to the muon $g-2$ is always smaller than the current muon $g-2$ anomaly (See the left panel of Fig.~\ref{fig4}).

In the large $n_c$ limit or in the limit of small curvature of the extra dimension, $k\ll M_5$, the leading contribution, $\Delta_{(1)}a_{\mu}^{\rm CW}$ to the muon $g-2$ is independent of the background geometry of extra dimension. Such a universal form of the leading contribution to the muon $g-2$  in the large $n_c$ limit encourages us to calculate the value in other types of background geometry.
Being a 5D theory in a warped geometry, the continuum clockwork theory is quite similar to the Randall-Sundrum model~\cite{Randall:1999ee}. But, the differences lie in the mass spectrum and the graviton couplings, which lead to quite different features generically in low-energy physics such as the muon $g-2$ as well as in collider physics.  The metric of RS models takes for $0\le y\le \pi R$
\begin{equation}
ds^2=e^{2ky}dx_{\mu}dx^{\mu}-dy^2
\end{equation}
but the 4D effective Planck mass is same as that of CW theory, Eq.~(\ref{planck}). The RS graviton has mass given by the zeros of the Bessel function of first kind $J_1(j_n)=0$ as 
\begin{equation}
M_{(n)}=kj_n\approx \left(n+\frac14\right)\pi k, \quad n=1,2,3,\cdots\,.
\end{equation}
and its 4D effective gravitational inverse-coupling
\begin{equation}
\Lambda_n\simeq\sqrt{\frac{M_5^3}{k}}\,.
\end{equation}
Similarly to the CW gravitons, the highest level that RS gravitons can be excited to is given by 
\begin{equation}
M_{(n_c)}\approx \alpha M_5 \quad {\rm or}\quad {n}_c\approx\frac{\alpha M_5}{\pi k}-\frac14\,,
\end{equation}
which is always smaller than the maximum value of the highest level, $n_c<{n_*}=n_c^{3/2}\sqrt{\pi}$,
at which $M_{(n_*)}\sim\Lambda_{n_*}$.
The one-loop contribution of RS gravitons to the muon $g-2$ anomaly is summed up to its allowed highest level $n_c$ to get~\footnote{The muon $g-2$ in the Randall-Sundrum model has been calculated with different cutoff, $n_c$ in~\cite{Park:2001uc,Kim:2001rc}.
The subleading part $\Delta_{(2)}a_{\mu}^{\rm RS}$ also could be shown scheme-independent by other regularization methods as well.} 
\begin{equation}
a_{\mu}^{\rm RS}=\sum_{n=1}^{n_c}a_{\mu}^{(n)}\approx\frac{5}{16\pi^3}\left(\frac{m}{M_5}\right)^2\left[\alpha-\frac{\pi k}{4M_5}\right]\,.\label{muong-2RS}
\end{equation}
We see that for large $n_c\gg1$ or $M_5\gg k$, both the clockwork and Randall-Sundrum models give at the leading order the same contribution of gravitons to the muon $g-2$, namely $\Delta_{(1)}a_{\mu}^{\rm RS}=\Delta_{(1)}a_{\mu}^{\rm CW}$. On the other hand the subleading but scheme-independent contribution of RS background is bigger than that of CW background. The CW contribution in the small curvature limit ($M_5\gg k$) is therefore smaller than the RS contribution. 

Finally we brief consider the massive graviton contributions to the muon $g-2$ in the large extra dimensions (LED), which has a flat 5D metric with the effective 4D Planck mass, given in terms of 5D intrinsic scale $M_5$ and the radius, $R$, of the 5th direction,
\begin{equation}
M_P^2=M_5^3\pi R\,.
\end{equation} 
Since the metric is flat and by the boundary condition of the graviton wave-function, $\psi(y+\pi R)=\pm\psi(y)$, the Kaluza-Klein mass spectrum of LED graviton takes with $n=1,2,3,\cdots$
\begin{equation}
M_{(n)}=\frac{n}{R}\,.
\end{equation}
Taking the highest level of LED gravitons $n_c=\alpha M_5R$, we find the total contribution of LED gravitons, 
\begin{equation}
a_{\mu}^{\rm LED}\approx \frac{5\alpha}{16\pi^3}\left(\frac{m}{M_5}\right)^2=4.5\times10^{-10}\alpha\left(\frac{0.5~{\rm TeV}}{M_5}\right)^2\,,
\end{equation}
which is nothing but the lowest upper bound of all warped extra dimensions as $\alpha$ is independent of the background geometry. 
For the next order contributions, if $k\sim M_5$, only a few massive gravitons are allowed, $n_c\sim {\cal O}(10)$,  and in this case RS gravitons always give bigger contributions to the muon $g-2$ than CW gravitons but smaller contributions, compared to LED gravitons, $a_{\mu}^{\rm CW}\lesssim a_{\mu}^{\rm RS}\lesssim a_{\mu}^{\rm LED}$, though all the extra dimensional models give similar contributions for the same intrinsic scale, $M_5$.  

\subsection{Results and Discussion}
We have calculated in the clockwork theory the contributions of massive gravitons to the anomalous magnetic moment of muon. We find that the clockwork gravitons do explain the current muon $g-2$ anomaly or $\Delta a_{\mu}\approx 28.8\times 10^{-10}$, if the intrinsic scale of the extra dimension $M_5\sim 0.2~{\rm TeV}$ with $k\lesssim 0.1~{\rm TeV}$, which corresponds to a quite large extra dimension, $l_5\gtrsim 10^{-7}\,{\rm m}$. For generic clockwork models, however, with $M_5\sim 1 - 100~{\rm TeV}$ and $M_5>  k$, the contributions of clockwork gravitons are too small to account for the current muon $g-2$ anomaly.  We also find that the lowest upper bound of the graviton contributions to the muon $g-2$ is given by the large extra dimension models among all the extra dimension models with the same 5D intrinsic scale $M_5$, if the UV sensitive contribution is same,  which might be useful in top-down model building. 
This property could be confirmed  for extra dimension models with more generic 5D metric background and KK graviton spectra that solve the gauge hierarchy problem~\cite{Choi:2017ncj}.

\subsection{Acknowledgements}
We thank A. Strumia for a useful comment. This research was supported by Basic Science Research Program through the National Research Foundation of Korea (NRF) funded by the Ministry of Education (NRF-2017R1D1A1B06033701) (DKH) and also by IBS under the project code, IBS-R018-D1 (DHK,CSS). This work was initiated at the CERN-CKC workshop ``What's going on at the weak scale?", Jeju Island, June 2017, and completed while one of the author (DKH) was enjoying the hospitality of the Aspen Center for Physics, which is supported by National Science Foundation grant PHY-1607611.   CSS is supported in part by the Ministry of Science, ICT\&Future Planning and by Gyeongsangbuk-Do and Pohang City.

\end{document}